\documentclass{JHEP3}

\pdfoutput=1

\usepackage{amsmath}
\usepackage{epsfig,multicol,bbm}
\usepackage{graphicx}
\usepackage{bm}
\usepackage{multirow}

\usepackage{subfigure}

\def\oas{O(\alpha_s)}

\newcommand{\orderalpha}{O(\alpha_s)}

\newcommand{\orderlambdao}{O(\lambda^0)}
\newcommand{\orderlambda}{O(\lambda)}
\newcommand{\orderlambdasqr}{O(\lambda^2)}

\newcommand{\gm}{\mbox{$\gamma_{\mu}$}}
\newcommand{\gn}{\mbox{$\gamma_{\nu}$}}
\newcommand{\al}{\alpha}
\newcommand{\ga}{\gamma}

\newcommand{\ep}{\epsilon^*}
\newcommand{\newcaption}[1]{\caption{#1}}

\def\be{\begin{equation}}
\def\ee{\end{equation}}
\def\ba{\begin{eqnarray}}
\def\ea{\end{eqnarray}}
\def\bfig{\begin{figure*}[htb]}
\def\efig{\end{figure*}}
\def\fig#1{Fig.\ \ref{#1}}
\def\sec#1{Sec.\ \ref{#1}}
\def\eqn#1{(\ref{#1})}
\newcommand{\nn}{\nonumber \\}

\def\nb{{\bar{n}}}
\def\lag{{\cal L}}
\def\J{{\cal J}}

\def\R{{(\bf R)}}

\def\Dslash{\fmslash{D}}
\def\nslash{\fmslash{n}}
\def\nbslash{\fmslash{\bar n}}
\def\pslash{\fmslash{p}}

\def\wnfun{W_n^{(\bf 3)}}
\def\wnbfun{W_{\bar n}^{(\bf 3)}}
\def\ynfun{Y_n^{(\bf 3)}}
\def\ynbfun{Y_{\bar n}^{(\bf 3)}}
\def\wnadj{W_n^{(\bf 8)}}
\def\wnbadj{W_{\bar n}^{(\bf 8)}}
\def\ynadj{Y_n^{(\bf 8)}}
\def\ynbadj{Y_{\bar n}^{(\bf 8)}}
\def\xnb{x_\nb}
\def\xn{x_n}
\def\xnbinf{x_\nb^\infty}
\def\xninf{x_n^\infty}
\def\xsninf{x_{n}^{\infty_s}}
\def\xsnbinf{x_{\nb}^{\infty_s}}
\def\xc{x}
\def\xsh{x_s}
\def\xshinf{\xsnbinf}
\def\xinftyc{x_c^\infty}
\def\xinftysh{\xshinf}

\def\deltascet{\delta^{(4)}_{\mbox{\tiny SCET}}}
\def\deltaqcd{\delta^{(4)}_{\mbox{\tiny QCD}}}
\def\deltaqcdbsg{\delta^{(4)}_{\mbox{\tiny QCD}}}

\def\ncol{$n$-collinear}
\def\nbcol{$\bar n$-collinear}
\def\amp{{\mathcal M}}

\makeatletter
\def\fmslash{\@ifnextchar[{\fmsl@sh}{\fmsl@sh[0mu]}}
\def\fmsl@sh[#1]#2{%
   \mathchoice
     {\@fmsl@sh\displaystyle{#1}{#2}}%
     {\@fmsl@sh\textstyle{#1}{#2}}%
     {\@fmsl@sh\scriptstyle{#1}{#2}}%
     {\@fmsl@sh\scriptscriptstyle{#1}{#2}}}
\def\@fmsl@sh#1#2#3{\m@th\ooalign{$\hfil#1\mkern#2/\hfil$\crcr$#1#3$}}

\makeatother

 \title{SCET, QCD and Wilson Lines}
 
 \author{
Simon M. Freedman and Michael Luke\\
Department of Physics, University of Toronto, Toronto, ON, Canada M5S1A7\\
E-mail: \email{sfreedma@physics.utoronto.ca}, \email{luke@physics.utoronto.ca}
}

\abstract{
Soft Collinear Effective Theory (SCET) is an effective field theory which describes the interactions
of low invariant mass jets which are highly boosted with respect to one another. In the standard
formulation of SCET, the effective Lagrangian for collinear fields is expanded in inverse powers of
the energy.  At leading order this leads to manifest decoupling of soft and collinear degrees of
freedom; however, subleading terms in the effective Lagrangian violate this manifest decoupling.  In
this paper we point out that the collinear expansion in the SCET Lagrangian is unnecessary, and that
the SCET Lagrangian may instead be written as  multiple decoupled copies of QCD.  The interactions
between the sectors in full QCD are reproduced in the effective theory by an external current
consisting of QCD fields coupled to Wilson lines. We illustrate this picture with two examples:
dijet production and $B\to X_s\gamma$.
}

\begin{document}
\maketitle

%==========Intro================

\section{Introduction}
\label{sec:intro}

Soft-Collinear Effective Theory (SCET) \cite{Bauer:2000ew, Bauer:2000yr,
Bauer:2001ct, Bauer:2001yt, Bauer:2002nz, Beneke:2002ph} describes the interaction of low invariant
mass jets of particles which are highly boosted with
respect to one another.  SCET is an expansion in inverse powers of the highly boosted energy.
At leading order in the SCET expansion, a field
redefinition may
be used to manifestly decouple the soft and collinear degrees of freedom from
one another at the operator level \cite{Bauer:2001yt}.   Interactions between different
soft and collinear sectors are reproduced in 
the currents of the effective theory by lightlike Wilson lines.
This simplification is the basis of factorization theorems in SCET, 
allowing differential cross sections to be written as convolutions of
independent soft and collinear pieces.
While factorization theorems have been well-studied using traditional QCD
approaches, the
manifest decoupling of soft and collinear pieces at the level of the Lagrangian
in SCET both 
dramatically simplifies the study of factorization theorems, and allows power
corrections in inverse energy to be studied
in a systematic way. 

In standard formulations of SCET \cite{Bauer:2000ew, Bauer:2000yr,
Bauer:2001ct, Bauer:2001yt, Bauer:2002nz, Beneke:2002ph,Beneke:2002ni}, there is an inherent
asymmetry in the treatment of soft and
collinear degrees of freedom.  While, for example, soft quark fields are
identical to four-component QCD quark fields,  collinear quark fields are
described by two-component spinors with complicated nonlocal interactions. 
On
general grounds, this asymmetry must be spurious:  QCD is Lorentz invariant, and
dimensional regularization is a Lorentz invariant regulator.   One may therefore
always boost to a 
reference frame in which the energy of the collinear fields is small, and
the collinear quark fields are 
described by four-component QCD fields.  Thus, the SCET description of collinear
fields must be equivalent
to that of full QCD.

This is not a new observation.  It was observed in \cite{Beneke:2002ph} that the
Feynman rules of collinear
SCET fields are equivalent to those of QCD in light-cone quantization
\cite{Bassetto:1984dq}, and
this equivalence has been used to simplify calculations in the collinear sector
of the theory \cite{Bauer:2000ew, Bauer:2000yr,
Becher:2006mr}.   In \cite{Bauer:2008qu}, it was formally proven at leading order
in power corrections that SCET is equivalent to multiple 
copies of QCD coupled to Wilson lines when the field redefinition of \cite{Bauer:2001yt}
is used to decouple soft from collinear fields. However, beyond leading order
the approach was less clear.

In this paper, we argue that 
this picture may be extended to all orders in the SCET expansion. 
We show that the soft and  collinear sectors
of SCET may individually be described by a separate copy of the full QCD
Lagrangian, and that these sectors are decoupled from one another to all orders
in the SCET expansion.  The interactions between the sectors in full QCD are
reproduced by the interactions between the individual sectors and the external current,
which consists of QCD fields coupled to Wilson lines.  In particular, soft-collinear 
mixing terms in the Lagrangian do not arise in the theory;
their effects are accounted for by subleading corrections to the external
current, whose form is similar to that of subleading twist shape
functions \cite{Ellis:1982cd,
Bauer:2001mh}.

In order to motivate this picture, we derive the subleading operators for two
specific phenomena; $e^+e^-\to$ dijet production and $B\to X_s\gamma$.  In
\sec{sec:scet} we review the standard derivation of SCET.  In \sec{sec:lodijet} we
present our approach for dijet production at leading order, while in \sec{sec:nlodijet} we derive the 
new subleading operators for dijet production.  In \sec{sec:bsg} we present a similar analysis for 
$B\to X_s\gamma$, and in \sec{sec:conc} we present our conclusions.  

%=============Label SCET=================

\section{Label SCET Formulation\label{sec:scet}}

In the approach to SCET introduced in \cite{Bauer:2000ew, Bauer:2000yr,Bauer:2001ct,
Bauer:2001yt, Bauer:2002nz}, collinear fields are described by effective two-component
spinors,   $\xi_{n,\tilde p}$, where $n$ denotes the (lightlike) direction of motion,
$\tilde p$ is a label which denotes the large
components of the collinear momentum,
\be
	\tilde p^\mu \equiv \nb\cdot p\, \frac{n^\mu}2+p_\perp^\mu
\ee
and the collinear quark momentum is $p^\mu=\tilde p^\mu+k^\mu$.    We will refer
to this approach as ``label SCET" 
to denote the removal of the large label momentum, and to distinguish it from
the approach of \cite{Beneke:2002ph,Beneke:2002ni}, 
in which label momentum was not removed, but the collinear quarks were still
treated as two-component spinors.
The SCET Lagrangian 
for the collinear quark field is obtained by integrating out the two small
components of the
field and expanding in powers of $\lambda^2\sim k^\mu/\nb\cdot  p$.   This
procedure results in the effective
Lagrangian for the $n$-collinear quark
\ba\label{scetoldlag}
	\lag_\xi=\lag^{(0)}_{\xi\xi}+\lag^{(1)}_{\xi\xi}+\lag^{(1)}_{\xi q}+\ldots,
\ea
where the superscript refers to the suppression in $\lambda$
\cite{Beneke:2002ph,Beneke:2002ni,Lee:2004ja,Pirjol:2002km, Bauer:2003mga,
Chay:2002vy}. 
The leading order term $\lag_{\xi\xi}^{(0)}$ is
\begin{eqnarray}\label{L0}
{\cal L}_{\xi\xi}^{(0)} 
&=&\sum_{\tilde p,\tilde p^\prime} \bar \xi_{n,\tilde p^\prime} \left[ i n\cdot
D
+ i \Dslash^\perp_n  W_n \frac{1}{\nb\cdot {\cal P}} W_n^\dagger 
  i \Dslash^\perp_n\right] \frac{\nbslash}{2} \xi_{n,\tilde p}  
\end{eqnarray}
where the covariant derivative $D_\mu=\partial_\mu-i g T^a A_\mu^a$, $A^\mu=A^\mu_s+A^\mu_n$
contains both soft and collinear gluons, 
$D_n^\mu$ only contains $n$-collinear gluons,  $W_n$ is a Wilson line built out
of collinear $A_n$ fields in the $\nb$ direction, 
and the ``label operator" $ {\cal P}^\mu$ pulls down the large label momentum of
the collinear fields.
The subleading operator $\lag_{\xi\xi}^{(1)}$ describes higher order corrections to the interactions 
in $\lag_{\xi\xi}^{(0)}$, while
the subleading operator $\lag_{\xi q}^{(1)}$ is the leading operator which couples collinear and soft quarks.
Performing the field
redefinitions on collinear quark and gluon fields \cite{Bauer:2001yt}
\ba\label{redef}
\xi_{n,\tilde p}(x)&=&\ynfun(x, \infty) \xi_{n,\tilde p}^{(0)}(x)\nn
A_{n,\tilde p}^{a\,\mu}(x)&=&\ynadj{}^{ab}(x, \infty) A_{n,\tilde p}^{(0)\, b\,\mu}(x),
\ea
where $Y_n^{\R}$ are Wilson lines built out of soft $A_s$ fields defined below,
it may be shown that all dependence on soft gluons disappears from the leading
order Lagrangian (\ref{L0}), so soft and collinear fields manifestly decouple
at leading order in SCET. The collinear and soft lightlike
Wilson lines in position space are defined as
\ba\label{wilsondef}
	W^\R_n(x,y)&=&P\exp\left(-ig\int_{0}^{\frac{n}2\cdot (y-x)}d s\,\nb\cdot
A^a_n(x+\nb s)T_{\bf R}^a\right)\nn
	Y^\R_n(x,y)&=&P\exp\left(-ig\int_{0}^{\frac{\nb}2\cdot (y-x)}d s\,n\cdot
A^a_s(x+n s)T_{\bf R}^a\right)
\ea
where ${\bf R}$ labels the $SU(3)$ representation.\footnote{In the SCET literature
\cite{Bauer:2001yt}, the fundamental Wilson lines ($\bf R=3$) are typically denoted by $W$ and $Y$,
and the adjoint Wilson lines ($\bf R=8$) are denoted by $\mathcal{W}^{ ab}$ and $ \mathcal{Y}^{ab}$,
where $a,b=1,\ldots,8$.}  Under a gauge transformation, the Wilson lines transform as
\begin{eqnarray}
W_n^\R(x,y)\to\mathcal{U}_c^\R(x) W_n^\R(x,y)\mathcal{U}_c^{\R\dagger}(y)\nn
Y_n^\R(x,y)\to\mathcal{U}_s^\R(x) Y_n^\R(x,y)\mathcal{U}_s^{\R\dagger}(y)
\end{eqnarray}
where $\mathcal{U}_{c,s}^{\R}$ is either a collinear or soft gauge transformation for representation
${\bf R}$.  Note that $W_n^{\R}(x,y)=W_n^{\R\dagger}(y,x)$, and similarly for $Y_n^{\R}$, and that
$W_n^\R(x,y)$ and $Y_n^\R(x,y)$ correspond to colour charge $\bf R$ propagating from $y$ to $x$. 
Also note that $W_n^\R$ and $Y_n^\R$ couple the $\nb$ and $n$ components of the corresponding
gluons, respectively; this notation is used to be consistent with the SCET literature.

At leading order, performing the field redefinitions (\ref{redef}), the current
for dijet production in the full theory
\be\label{qcddijetcurrent}
\J_2^{\rm QCD}= e^{-i Q\cdot x}\; \bar \psi(x)\Gamma\psi(x)
\ee
(where $\Gamma$ is an arbitrary Dirac structure and $Q$ is the external momentum) may be written in
the factorized form in the effective theory  
\be\label{usualdijetcurrent}
\tilde\J^{(0)}_2= e^{-i Q\cdot x} \;\bar\xi^{(0)}_{n,\tilde p_1}(x) \wnfun(x, \infty) 
\ynfun{}^\dagger(x, \infty) \Gamma \ynbfun(x, \infty)
\wnbfun{}^\dagger(x, \infty) \xi^{(0)}_{\nb, \tilde p_2}(x)
\ee
where the $W_\nb$ and $Y_\nb$'s are lightlike Wilson lines defined analogously to (\ref{wilsondef}). 
Label momentum conservation is enforced at each vertex. The collinear Wilson line $\wnfun$ arises
from integrating out the interactions of \ncol\  fields with \nbcol\ fields and similarly for
$\wnbfun$. Each sector is therefore decoupled at leading order and described by QCD fields coupled
to Wilson lines.

In this form it is manifest that all interactions between the
different sectors occur via Wilson lines, as was formally shown in
\cite{Bauer:2008qu}.   The redefined quark fields $\xi^{(0)}$ do not transform
under soft gauge transformations, so the soft fields only couple to the Wilson
line $Y$.  Physically, this corresponds to the fact that soft fields cannot
deflect the worldline of a highly energetic quark, and so they only see the
direction and gauge charge of the collinear degrees of freedom (much the same
way that in Heavy Quark Effective Theory \cite{Manohar:2000dt}, soft degrees of freedom
only see the velocity and gauge charge of heavy quarks).  Similarly, in a frame
in which the  $n$-collinear quark fields are soft, the soft and \nbcol\ fields
are recoiling in the opposite direction; thus the $n$-collinear quark fields can
only resolve the total gauge charge of the combined soft and \nbcol\ fields via
the Wilson line $W_{n}$ (and similarly for the \nbcol\ fields).

At higher orders in the expansion, however, label SCET looks more complicated, and the
operator decoupling is no longer manifest.  In particular, interactions such as
$\lag_{\xi q}^{(1)}$, in which soft and collinear sectors couple directly
instead of via Wilson lines \cite{Bauer:2003mga}, make the extension of the
arguments in \cite{Bauer:2008qu} to higher orders unclear.  In the next section
we show how the leading order picture can be easily extended, by reformulating
the theory using QCD fields.

Another formulation of SCET \cite{Beneke:2002ph, Beneke:2002ni} replaces the removal of label
momentum with a multipole expansion in soft position.  Our formulation of SCET more closely
resembles this formulation than label SCET.  However, we will diverge from the \cite{Beneke:2002ph,
Beneke:2002ni} treatment of collinear quarks, which are two-component spinors giving mixed
collinear-soft Lagrangian terms at subleading orders similar to label SCET.  Also, the non-Abelian
nature of SCET requires the introduction of a Wilson line $R(x)$ \cite{Beneke:2002ni}.  Without the
$R$ Wilson line, soft transformations of collinear fields gives higher order in $\lambda$ pieces due
to the soft and collinear fields being at different positions.  The $R$ Wilson line redefines the
collinear fields so they transform homogeneously in $\lambda$ under soft transformations.  However,
after the field redefinition \eqn{redef}, collinear fields no longer transform under the soft gauge
group, and the $R$ Wilson line is not needed.  In our formulation, soft and collinear fields are
decoupled and each sector does not transform under the other so the $R$ Wilson line will be
unneeded.

%============SCET=QCD coupled WL====================

\section{SCET as QCD Fields Coupled to Wilson Lines\label{sec:dijet}}

Despite the complexity of the leading order \ncol\ Lagrangian (\ref{L0})
and the corresponding Feynman rules,
it is equivalent to the QCD Lagrangian \cite{Bauer:2008qu, Beneke:2002ph}.  This is not unexpected:  
as long as one
is just describing soft fields or collinear fields in one direction,
there is no Lorentz-invariant expansion parameter, and one could just as easily
work in a frame where the energy is small, in which case it is obvious that
there is no effective field theory description and QCD is the appropriate
theory.  The large boost of a collinear quark only has 
physical meaning when it is coupled to fields with large relative momentum via
an external current, such as in $e^+ e^-\to q\bar q X$ 
or $B\to X_s\gamma$.
The purpose of SCET is to describe the interactions in such situations between
fields whose relative momentum is greater than the cutoff of the theory.

We therefore begin with the starting point that in the absence of an external
current, each sector (collinear in each relevant direction and soft) can be
described by $\lag_{\rm QCD}$, since QCD is Lorentz invariant. Therefore, the
all-orders SCET Lagrangian is
\begin{equation}\label{scetlag}
	\lag_{\rm SCET}=\sum_{i=s,n_j}\lag_{\rm QCD}^i,
\end{equation}
where $j$ runs over all relevant collinear directions.  $\lag_{\rm SCET}$ then
 consists of a separate copy of the QCD Lagrangian for each sector, 
each with a separate gauge symmetry.
All interactions between the different sectors will be described by the external current,
which for dijet production takes the form
\begin{equation}\label{scetcurrent}
\J_2^{\rm SCET}=e^{-i\frac{Q}2 x\cdot(n+\nb)} \left[C_2^{(0)} O_2^{(0)}+\frac1Q\sum_i
C_2^{(1i)} O_2^{(1i)}+\orderlambdasqr\right]
\end{equation}
where $O_2^{(0)}$ is $O(1)$ and the $O_2^{(1i)}$'s are $\orderlambda$, and we have pulled out the 
phase corresponding to the momentum of the external current. This is the only place the $\lambda$ 
expansion enters in this formulation of SCET.

As discussed in the previous section, fields in one sector only resolve the direction and
colour charge of fields in other sectors; hence, 
the sectors can only interact with each other via Wilson lines. The current
$\J_2^{\rm SCET}$ therefore decouples into separately $SU(3)$-invariant pieces
representing each sector, each of which describes QCD fields coupled to Wilson
lines.  At leading order the current $O_2^{(0)}$ is equivalent to the usual leading order
SCET current (\ref{usualdijetcurrent}).  The subleading operators $O_2^{(1i)}$
are
constructed from Wilson lines with derivative insertions, in a similar manner as
higher twist corrections to light-cone distribution functions
\cite{Ellis:1982cd,Bauer:2001mh}.

We will show that we can do this to subleading order for dijet and
heavy-to-light currents with nonlocal operators.  It will prove unnecessary to
introduce large label momenta, since these are frame-dependent. Instead, we
follow \cite{Bauer:2000ew} and \cite{Beneke:2002ph, Beneke:2002ni} and implement the
appropriate multipole expansion through the coordinate dependence of the
currents.   We first work out the leading order operators to illustrate our
picture in the next section, and then describe the subleading corrections.  We
demonstrate that all such corrections may be accounted for by subleading
corrections to the current, rather than direct interactions between the
different sectors (such as the collinear-soft quark interaction term ${\cal
L}_{\xi q}$).   This is the principal result of this paper.

%===============LO Dijet====================

\subsection{Dijet Production at leading order\label{sec:lodijet}}

\begin{figure}[thb]
   \centering
   \subfigure[\label{fig:dijetqcd1}]{  
\includegraphics[height=0.1\textheight]{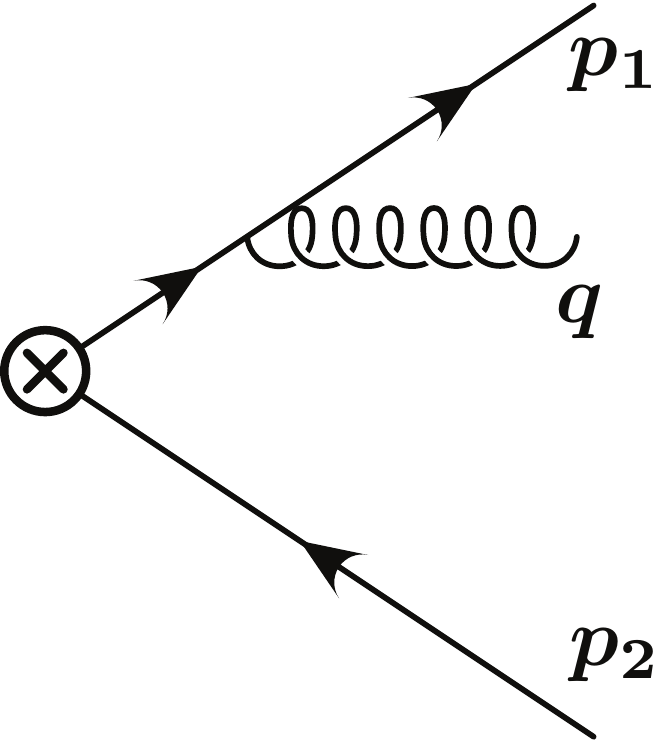}
}\hspace{0.1\textwidth}
   \subfigure[\label{fig:dijetqcd2}]{\includegraphics[height=0.1\textheight]
   {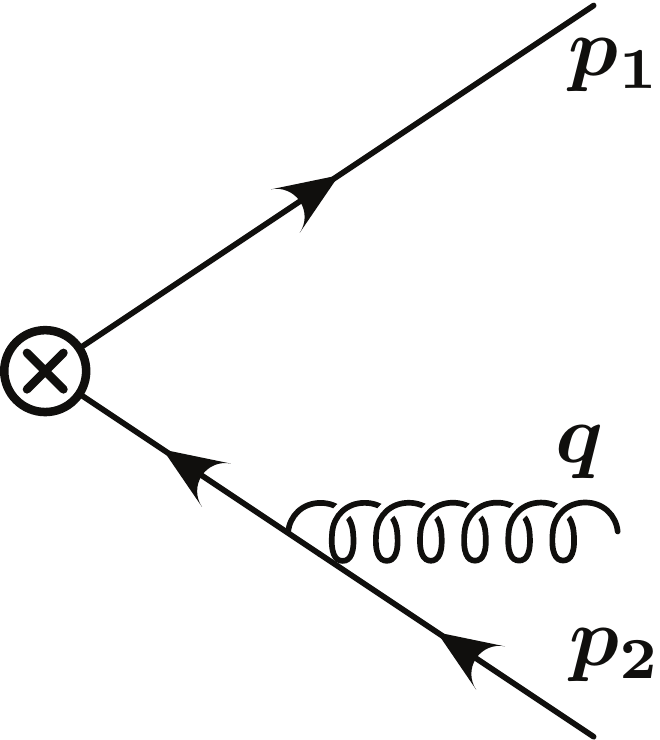}}
   \caption{QCD vertex for dijet production.  In this and all other figures, the quark,
   antiquark and gluon momenta are denoted $p_1$, $p_2$ and $q$.\label{fig:dijetqcd}}
\end{figure}

Consider the process $e^+e^-\to q\bar q X$, which contributes to dijet production.  The external 
current carries momentum
\be
	Q_{e^+e^-}^\mu=Q\frac{n^\mu}2+Q\frac{\nb^\mu}2,
\ee
where $Q$ is large compared to the invariant mass of the jets.  The $O(\alpha_s)$ 
graphs contributing to this process in QCD are shown in \fig{fig:dijetqcd}.

%%%%%%%%%%%%%%%%%%%%%%
%%%%%%%%%%%%%%%%%%%%%%

The SCET expansion of a given graph depends on the relative scaling of the
momenta:   $n$-collinear momenta scale like $p_n\sim Q(\lambda^2,1,\lambda)$,
\nbcol\ momenta like $p_{\bar n}\sim Q(1,\lambda^2,\lambda)$ and soft momenta
like $p_s\sim Q(\lambda^2, \lambda^2, \lambda^2)$\footnote{We use light-cone
coordinates, where $p^\mu=(p\cdot n, p\cdot\nb, {\vec p_\perp})$ and $n\cdot\nb=2$.}.  To match amplitudes
onto SCET, we expand the relevant graphs with the appropriate scalings in powers
of $\lambda$, including the various energy-momentum conserving delta functions. 
In particular, the full theory energy-momentum conserving delta function is 
\ba\label{qcdmm}
	\deltaqcd(Q; p)\equiv\delta^{(4)}(Q^\mu_{e^+e^-}-p^\mu)&=&2\,\delta\left(Q-p\cdot
n\right)\delta\left(Q-p\cdot\nb\right)\delta^{(2)}\left(\vec p_\perp\right)
\ea
where $p^\mu$ is the four-momentum of the final state.  Splitting $p^\mu$ into
$n$-collinear, \nbcol\ and soft momenta,
\begin{equation}
p^\mu=p_n^\mu+p_{\bar n}^\mu+p_s^\mu
\end{equation}
 and expanding in powers of $\lambda$ gives at leading order the SCET
energy-momentum conserving $\delta$ function
\ba\label{scetdelta}
	\deltaqcd(Q; p)&=&\deltascet(Q; p_n, p_\nb)+p_{s\perp}^\mu{\frac{\partial}{\partial p_{n\perp}^\mu}}
	\deltascet(Q; p_n, p_\nb)+\orderlambdasqr
\ea
where
\be\label{deltalo}
	\deltascet(Q; p_n, p_\nb)=2\,\delta\left(Q-p_n\cdot
\nb\right)\delta\left(Q-p_\nb\cdot n\right)\delta^{(2)}\left(\vec p_{n\perp}+\vec p_{\nb\perp}\right)
\ee
and the first term in \eqn{scetdelta} is $O(1)$ and the second is $\orderlambda$.
Note that soft momenta are unconstrained by overall energy-momentum conservation
in the effective theory.
This expansion differs from the label SCET derivation, which replaces
\eqn{deltalo} with label conservation $\delta_{\tilde p,\tilde p'}$, and which
conserves momentum exactly in the effective theory.  Higher order terms in the
expansion of (\ref{qcdmm}) are accounted for by higher order corrections in
SCET.

The expansion \eqn{scetdelta} can be understood in calculations as expanding QCD
phase-space in SCET momentum, where subleading phase-space  effects are
incorporated into the subleading current through the higher multipole moments. Such was the case when 
considering phase-space of jets at $\orderlambdao$ \cite{Cheung:2009sg,Ellis:2010rw}.   

\begin{figure}[tbh] 
   \centering
  \includegraphics[height=0.15\textheight]{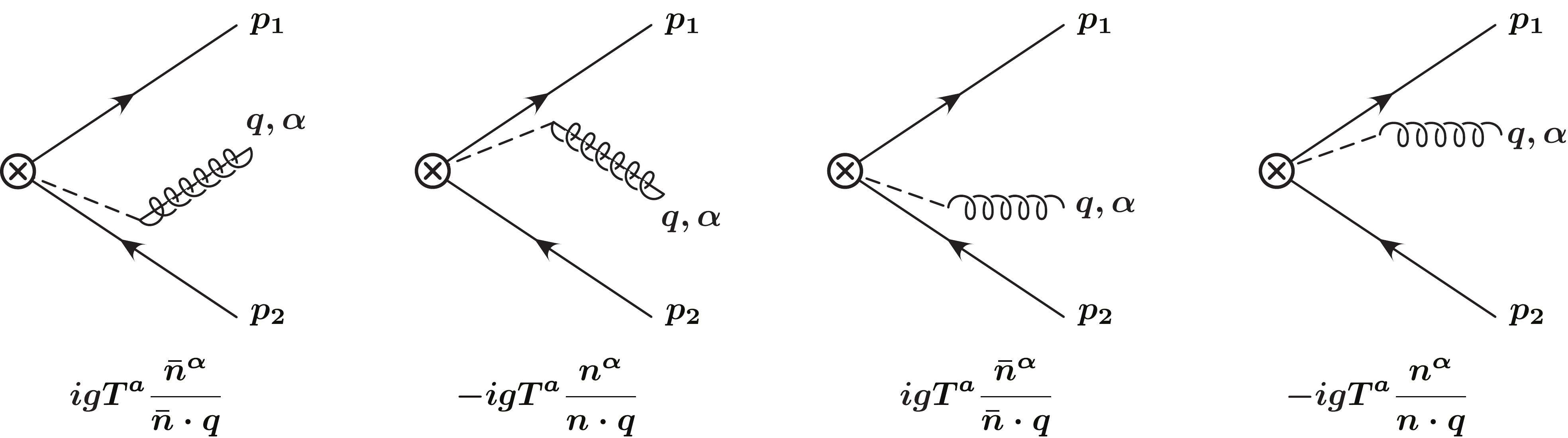}
   \caption{One-gluon Feynman rules for $O_2^{(0)}$.  The SCET energy-momentum
conserving delta function $\delta_{\rm SCET}$ has been omitted but is
implied.   Springs are soft gluons, springs with lines are  collinear
gluons,  and solid lines are fermions.  Lines angled up are \ncol, angled down
are \nbcol, and horizontal are soft.  The dashed lines represent the emission
from a Wilson line.\label{o2frules}}
\end{figure}

\begin{figure}[t] 
   \centering
   \subfigure[\label{fig:o2n}
\ncol]{\includegraphics[height=0.12\textheight]{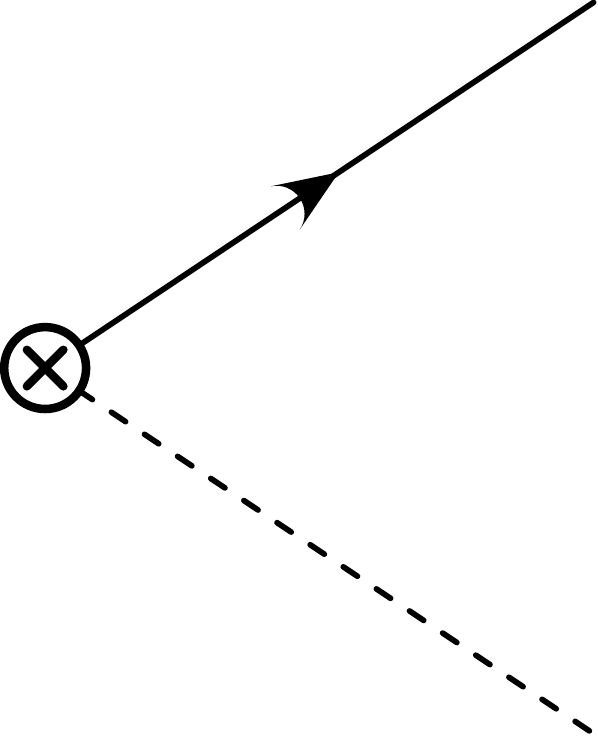}}\hspace{0.05
\textheight}
   \subfigure[\label{fig:o2s}
soft]{\includegraphics[height=0.12\textheight]{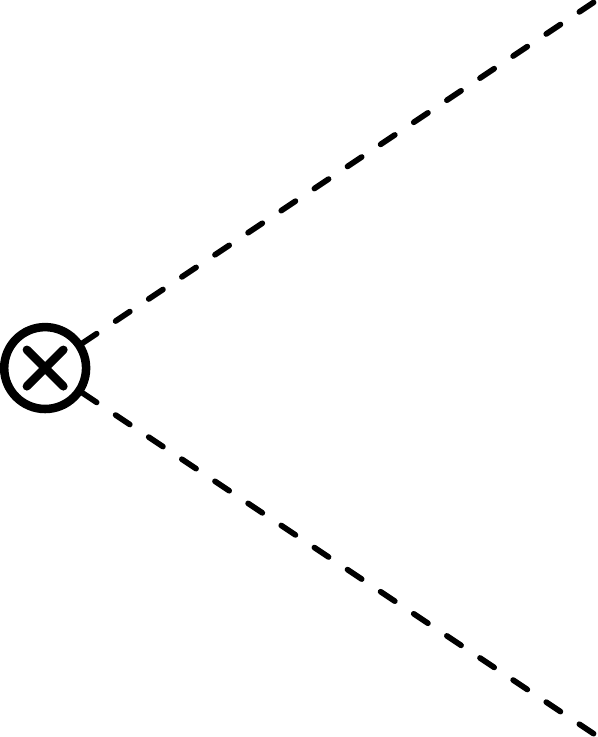}}\hspace{0.05 
\textheight}
   \subfigure[\label{fig:o2nb}
\nbcol]{\includegraphics[height=0.12\textheight]{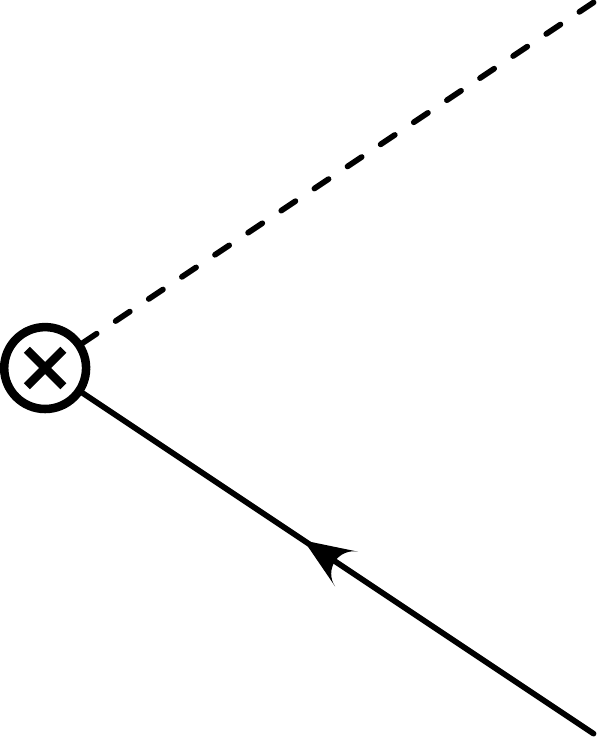}}\hspace{0.05
\textheight}   
   \caption{Physical picture of $O_2^{(0)}$ as seen in each of the three sectors.  The dashed lines 
represent Wilson lines and the solid lines represent fermions.\label{fig:o20}}
\end{figure}

We can write the external production current \eqn{scetcurrent} in
terms of four-component QCD spinors $\psi_n$ and $\psi_\nb$.  The leading
order operator is
\ba\label{o20}
	O_2^{(0)}(x)&=&\left[\bar\psi_n(\xnb)P_{\bar n}\Gamma  \wnfun(\xnb, \xnbinf)\right]\left[\ynfun(\xsninf, 0)\ynbfun(0,
\xsnbinf)\right]\left[ \wnbfun(\xninf, \xn)P_\nb\psi_\nb(\xn)\right]\nn
\ea
with $C_2^{(0)}= 1 +O(\alpha_s)$.  This has a similar form as \eqn{usualdijetcurrent}, with the difference that 
the collinear fields are four-component spinors, 
and the positions of the fields 
\ba\label{position}
	&\xn=(0,x\cdot\nb,\vec{x}_\perp)\quad    &
\xninf=(0,\infty,\vec{x}_\perp)\nn
	&\xnb=(x\cdot n, 0,\vec{x}_\perp)\quad	& \xnbinf=(\infty, 0,
\vec{x}_\perp)\nn
	&\xsninf=(0,\infty,0) \quad & \xsnbinf=(\infty,0,0),
\ea
are chosen to obtain the correct momentum conservation \eqn{deltalo}.  Note the coordinate $\xnb$ 
conserves $p\cdot \nb$ momentum, and similarly for $\xn$.   We have
also defined the usual projectors
\ba\label{projectors}
	P_n=\frac{\fmslash{n}\fmslash{\nb}}{4}\quad\quad\quad
P_\nb=\frac{\fmslash{\nb}\fmslash{n}}{4}
\ea
so at leading order in $\lambda$ the external current \eqn{o20} only couples to
the large components of the external quark spinors.  However, the collinear quark fields evolve via 
QCD, which couples all four components of the field.  The one-gluon Feynman rules
for $O_2^{(0)}$ are shown in \fig{o2frules}.  
%%%%%%%%%%%%%%%%%%%%
%%%%%%%%%%%%%%%%%%%%
The terms in each square bracket of \eqn{o20} each transform under a separate
$SU(3)$ symmetry, corresponding to the various sectors of the theory\footnote{We
ignore possible gauge transformations at $\infty$ since we use covariant gauge,
which is ``regular''. For complications that arise in ``singular'' gauges, see
\cite{Idilbi:2010im}. In our formulation, the necessary extra Wilson lines should occur naturally 
in the matching.} and represents a different decoupled sector giving the physical
picture of \fig{fig:o20}, which we explain below.

%%%%%%%%%%%%%%%%%%%%%%
%%%%%%%%%%%%%%%%%%%%%%

It is straightforward to show that the one-gluon matrix element of $O_2^{(0)}$
reproduces the QCD amplitude at leading order in $\lambda$.
The one-gluon amplitudes in \fig{fig:dijetqcd} in QCD are
\begin{eqnarray}\label{qcd1amp}
	i\amp_a=-igT^a\bar u(p_1)\frac{2p_1^\al+\ga^\al\fmslash{q}}{2p_1\cdot
q}\Gamma v(p_2)\ep_\al(q)\deltaqcd(Q; p_1+p_2+q)
\end{eqnarray}
and
\begin{eqnarray}\label{qcd2amp}
	i\amp_b&=&ig T^a\bar
u(p_1)\Gamma\frac{2p_2^\al+\fmslash{q}\ga^\al}{2p_2\cdot
q}v(p_2)\ep_\al(q)\deltaqcd(Q; p_1+p_2+q)
\end{eqnarray}
where $\Gamma$ is the Dirac structure of the external current.
The corresponding leading order contributions in SCET comes from an \ncol\
quark, \nbcol\ antiquark, and a gluon which is either soft or collinear, each of
which
gives a different result in SCET.

We first consider the case in which the \ncol\ quark emits an \ncol\ gluon.  
Using the Dirac Equation $\pslash_nu(p_n)=0$ to write
\ba\label{spinorexp}
	u(p_n)&=&\left(1+\frac{\fmslash{p}_{n,\perp}}{\nb\cdot
p_n}\frac{\fmslash{\nb}}2\right)P_n u(p_n)
\ea
and similarly for $v(p_{\bar n})$, it is straightforward to show that
\begin{eqnarray}\label{ncolexpandLO}
\bar u(p_1)\left(2 p_1^\alpha+\ga^\al\fmslash{q}\right)\Gamma v(p_2)=\bar u(p_1)
\left(2 p_1^\alpha+\ga^\al\fmslash{q}\right)P_{\bar
n}\Gamma P_{\bar n}v(p_2) +\orderlambda
\end{eqnarray}
so we can expand \eqn{qcd1amp} as
\begin{eqnarray}\label{ncolexp}
	i\amp_{an}&=&-igT^a\bar u(p_1)\frac{2p_1^\al+\ga^\al\fmslash{q}}{2p_1\cdot
q}P_{\bar n}\Gamma P_{\bar n} v(p_2)\ep_\al(q)\deltascet(Q; p_1+q, p_2)+\orderlambda.
\end{eqnarray}
With the projectors $P_{\bar n}$ now surrounding the Dirac structure $\Gamma$,
this is precisely the amplitude in the effective theory for a $q$-$\bar q$ pair
to be produced by $O_2^{(0)}$, followed by the emission of an \ncol\ gluon from the
\ncol\ quark through the usual QCD vertex.  It is useful to compare this with
the expression for the same graph in label SCET:
\ba\label{lscetexp}
i\amp_{an}^\prime&=&-igT^a \bar \xi_{n,\tilde p_1}(k_1) \left(n^\alpha+{\ga^\al_\perp
\fmslash{p}_{1\perp}\over\nb\cdot p_1}+{(\fmslash{p}_{1\perp}+\fmslash{q}_\perp)\ga^\al_\perp\over 
\nb\cdot (p_1+q)}+{(\fmslash{p}_{1\perp}+\fmslash{q}_\perp)\fmslash{p}_{1\perp}\over \nb
\cdot p_1\; \nb \cdot(p_1+q)}\nb^\alpha\right)\frac{\fmslash{\nb}}2\nn
&\times&\left(\frac{\fmslash{n}}2{\nb\cdot(p_1+q)\over n\cdot(k_1+k_3)\; \nb \cdot(p_1+q)+
(p_{1\perp}+q_\perp)^2}\right)\Gamma\xi_{\nb, \tilde p_2}(k_2)\ep_{n,\tilde q\,\al}(k_3)\nn
&\times&\delta_{\nb\cdot(\tilde p_1+\tilde q), Q}\delta_{n\cdot\tilde p_2, Q}
\delta_{0,\tilde p_{1\perp}+\tilde p_{2\perp}+\tilde q_{\perp}}\delta^{(4)}(k_1+k_2+k_3)
\ea
where the first factor in parentheses is the collinear quark - collinear quark - collinear gluon
vertex,  the second is the collinear quark propagator in label SCET, and the $\xi$ fields are
two-component spinors. Some straightforward Dirac algebra shows that this is indeed equivalent to
the expression \eqn{ncolexp}; however, the more complicated Feynman rules of label SCET, arising
from the fact that the collinear spinors are 2-component objects rather than 4-component spinors
obeying \eqn{spinorexp}, makes the intermediate expression considerably more complicated.

Expanding the amplitude in which an \ncol\ gluon is emitted from an \nbcol\
antiquark, (\ref{qcd2amp}),  in powers of $\lambda$ gives
\begin{eqnarray}\label{loex}
	i\amp_{bn}=ig T^a\frac{\nb^\al}{\nb\cdot q}\bar{u}(p_1)P_\nb \Gamma
P_\nb v(p_2)\ep_\al(q)\deltascet(Q; p_1+q, p_2)+\orderlambda
\end{eqnarray}
where we have used the expansions
\ba\label{exp}
	2p_2\cdot q&=&(p_2\cdot n)(q\cdot\nb)+\orderlambdasqr\nn
	2p_2\cdot\ep(q)&=&(p_2\cdot n)(\ep(q)\cdot\nb)+\orderlambdasqr.
\ea
In SCET, the \ncol\ quark does not couple to the \nbcol\ antiquark directly, but
rather to the Wilson line $\wnfun (\xnb, \xnbinf)$ in $O_2^{(0)}$, and this amplitude is
reproduced in the effective theory by the graph in which the Wilson line emits
the \ncol\ quark.  The interactions \eqn{ncolexp} and \eqn{loex} of the \ncol\ gluon is represented
in \fig{fig:o2n} by a QCD quark field in the $n$ direction and a Wilson line in the $\nb$ direction.

Similarly,  the amplitudes in \fig{fig:dijetqcd} are reproduced for \nbcol\ gluons in
SCET by a gluon emitted from a semi-infinite \ncol\ Wilson line $\wnbfun(\xn,
\xninf)$, and the usual QCD Feynman rules for gluon emission, respectively.  The \nbcol\ gluon 
interaction is represented in \fig{fig:o2nb}.

Finally, the amplitude for soft gluon emission from the quark and antiquark
lines is obtained by expanding the sum of the two previous graphs for soft gluon
momentum,
\ba
	i\amp_{s}=-ig T^a\left(\frac{n^\al}{n\cdot q}-
\frac{\nb^\al}{\nb\cdot q}\right)\bar{u}(p_1)P_n \Gamma P_\nb
v(p_2)\ep_\al(q)\deltascet(Q; p_1, p_2)+\orderlambda,
\ea
which is the amplitude for gluon emission from a fundamental and anti-fundamental Wilson
line, $\ynfun(\xsninf,0)$ and $\ynbfun(0,\xsnbinf)$ respectively, represented in \fig{fig:o2s}.

Thus, we have shown that $O_2^{(0)}$ as defined in \eqn{o20} reproduces the leading
order QCD $q\bar q g$ production amplitudes.  In the next section, we will show how $\orderlambda$ 
operators arise as generalizations of $O_2^{(0)}$ and the physical picture of \fig{fig:o20}. 

%====================Subleading Dijet==========================

\subsection{Subleading Corrections to Dijet Production\label{sec:nlodijet}}

\begin{figure*}[tb]
   \centering
   \includegraphics[width=0.9\textwidth]{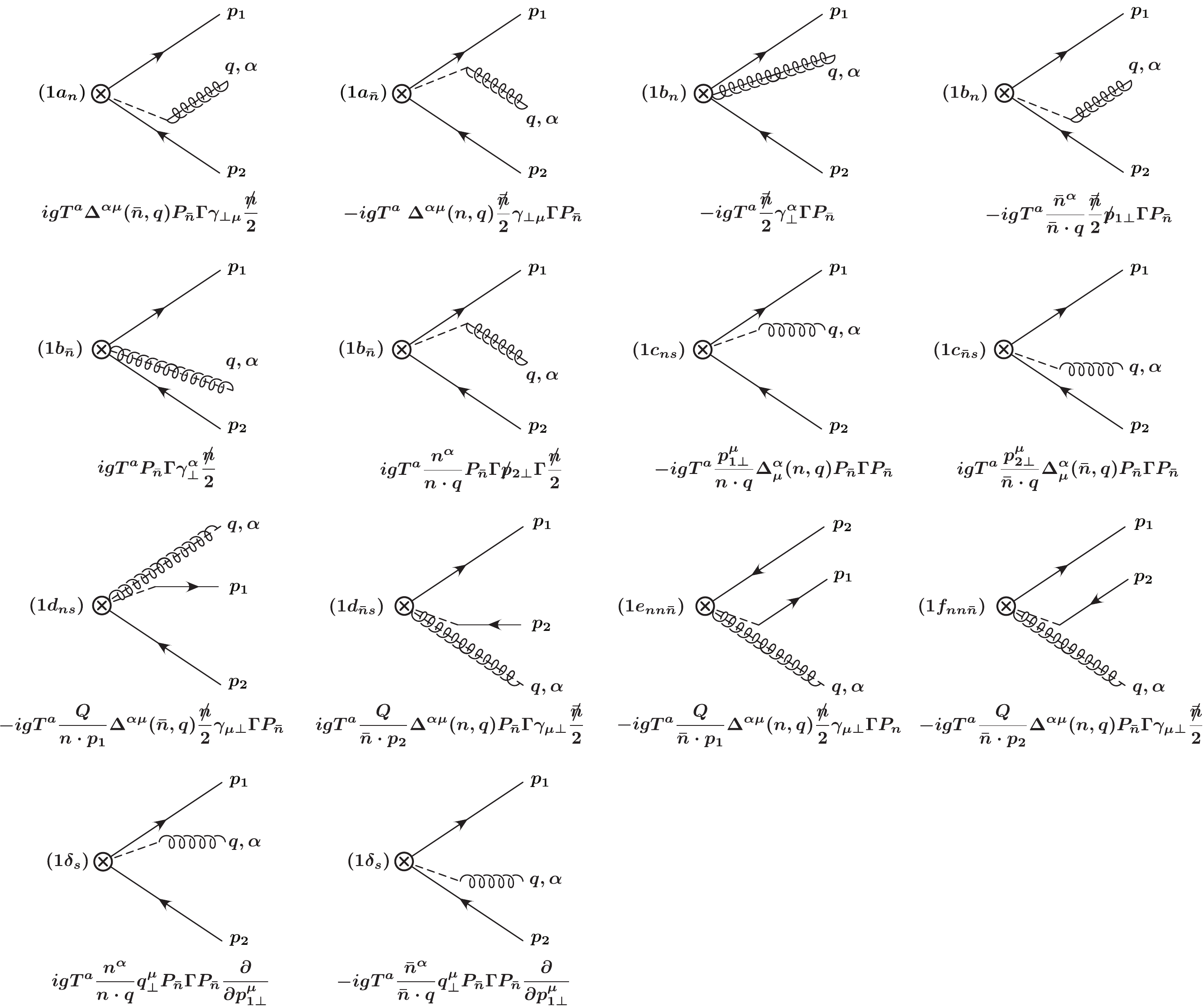} 
   \newcaption{One-gluon Feynman rules for NLO dijet operators $O_2^{(1i)}$.  The notation is the 
   same as \protect\fig{o2frules}.  The rule for $O_2^{(1b_n)}$ has been split into two diagrams, 
   depending on whether the gluon is emitted from the vertex or the Wilson line.\label{subleadingO2rules}}
\end{figure*}

At leading order, the external current is written as a product of QCD fields
coupled to Wilson lines.  Higher order corrections to the current are therefore
expected to have the same structure, but with  insertions of derivatives and
additional fields, in the same way that subleading twist shape functions and
parton distributions are related to the leading order operators \cite{Ellis:1982cd,
Bauer:2001mh}.  Defining the external current to subleading order in \eqn{scetcurrent}
where the $O_2^{(1i)}$'s are $\orderlambda$,  
it is straightforward to determine the required operators $O_2^{(1i)}$ and coefficient functions
$C_2^{(1i)}$ by
carrying out the expansion of the previous section to higher orders in
$\lambda$.  Starting with the emission of an \ncol\ gluon, we can expand the QCD
amplitudes (\ref{qcd1amp}) and (\ref{qcd2amp}) to $\orderlambda$:
\ba\label{ncolexpAsub}
	i\amp_{(a,b)n}=-igT^a\bar u(p_1)\Gamma^\al_{(a,b)n}
v(p_2)\ep_\al(q)\deltascet(Q; p_1+q, p_2)+\orderlambdasqr
\ea
where
\ba\label{aampexp}
\Gamma^\al_{an}&=&{\left(2 p_1^\alpha+\ga^\al\fmslash{q}\right)\over2 p_1\cdot q}P_{\bar n}
\Gamma P_{\bar n}-{\Delta^{\alpha\mu}(\bar n, p_1)\over
\nb\cdot(p_1+q)} \frac{\nbslash}2\gamma_{\mu\perp}\Gamma P_{\bar n} +{\left(2 p_1^\alpha+\ga^\alpha
\fmslash{q}\right)\over 2 p_1\cdot q}{\nbslash\over 2}
{\left(\fmslash{p}_{1\perp}+\fmslash{q}\right)\over \bar
n\cdot\left(p_1+q\right)} \Gamma P_{\bar n} \nn
\ea
and
\ba\label{bampexp}
	\Gamma^\al_{bn}&=&-\frac{\nb^\al}{\nb\cdot q}P_\nb \Gamma P_\nb-\frac{\nb^\al}{\nb\cdot
q}\frac{\fmslash{\nb}}2\frac{\fmslash{p}_{1\perp}}{\nb\cdot p_1}\Gamma P_\nb+\frac{1}{Q}\Delta^{\al\mu}(\nb,q)
P_{\bar n}\Gamma\gm_\perp\frac{\fmslash{n}}{2}
\ea
where we have defined
\be\label{Delta}
	\Delta^{\al\mu}(\bar n, p)=g^{\al\mu}-\frac{\bar n^\al p^\mu}{\bar n\cdot p}
\ee
and we have used the expansion
\be
	\fmslash{q}\fmslash{\epsilon}^*(q)=(\nb\cdot
q)\frac{\fmslash{n}}2\gm_\perp\Delta^{\al\mu}(\nb,q)\ep_\al(q)+\orderlambdasqr
\ee
as well as the spinor expansion \eqn{spinorexp}.  
The sum of the graphs is
\ba\label{nampexp}
\Gamma^\al_{an}+\Gamma^\al_{bn}&=&{\left(2
p_1^\alpha+\ga^\al\fmslash{q}\right)\over2 p_1\cdot q}P_{\bar n} \Gamma P_{\bar
n}-\frac{\nb^\al}{\nb\cdot q}P_\nb \Gamma P_\nb\nn
&+&\frac{1}{Q}\left[\Delta^{\al\mu}(\nb,q)P_{\bar n}\Gamma\gm_\perp\frac{\fmslash{n}}{2}+
\frac{\fmslash{\nb}}{2}\left(\gamma^\alpha_\perp+\frac{\nb^\alpha}{\nb\cdot
q}\fmslash{p}_{1\perp}\right)\Gamma P_\nb\right.\nn
&+&\left.{\left(2 p_1^\alpha+\ga^\al\fmslash{q}\right)\over 2 p_1\cdot q}{\nbslash\over 2}
\left(\fmslash{p}_{1\perp}+\fmslash{q}\right) \Gamma P_{\bar n}\right].
\ea
The first two terms of
\eqn{nampexp} are $O(1)$, while the remaining terms are
$\orderlambda$, and are reproduced in the effective theory by
the operators
\ba
	O^{(1a_n)}_{2}&=&\left[\bar\psi_n(\xnb)P_\nb\Gamma
i\fmslash{D}_{\perp}(\xnb)
\wnfun(\xnb,\xnbinf)\right]\left[ \ynfun(\xsninf,0) \ynbfun(0,
\xsnbinf)\right]\nn
&&\qquad\times\left[\wnbfun(\xninf,
\xn)\frac{\fmslash{n}}{2}\psi_\nb(\xn)\right]\nn
	O_2^{(1b_n)}&=&
\left[\bar\psi_n(\xnb)\frac{\fmslash{\nb}}2i\overleftarrow{\fmslash{D}}_{\perp}(
\xnb)\Gamma \wnfun(\xnb, \xnbinf)\right]\left[ \ynfun(\xsninf,0) \ynbfun(0,
\xsnbinf)\right]\nn
&&\qquad\times\left[\wnbfun(\xninf,
\xn)P_\nb\psi_\nb(\xn)\right]
\ea
where the covariant derivatives are defined in the usual way
\begin{equation}
D^\mu \psi_{n,\nb,s}=\partial^\mu \psi_{n,\nb,s}-ig T^a A^{\mu a}_{n,\nb,s} \psi_{n,\nb,s}
\end{equation}
to only couple the corresponding gluon fields to \ncol, \nbcol, and soft quarks, respectively.
The one-gluon Feynman rules for these operators are given in \fig{subleadingO2rules}.  The
last term in \eqn{nampexp} corresponds in the effective theory to a gluon
emitted from the $n$-collinear quark leg via $\lag_{\rm QCD}$ after
the insertion of the subleading operator $O_2^{(1b_n)}$.  From \eqn{nampexp}, we find
$C_2^{(1a_n)}=-1+\oas$ and $C_2^{(1b_n)}=1+\oas$.  

We can perform a similar expansion for soft gluon emission.  Expanding the
amplitude \eqn{qcd1amp} for soft momentum $q^\mu$, including the multipole expansion \eqn{scetdelta}, gives
\ba\label{softexp}
i\amp_{as}&=&-ig T^a\left[\bar u(p_1)P_{\bar n}\Gamma P_{\bar
n} v(p_2)\left({n^\alpha\over n\cdot
q}\left(1+q^\mu_\perp{\partial\over\partial p_{1\perp}^\mu}\right)\right.+
\frac{2p_{1\perp\mu}}{Q\; n\cdot q}\Delta^{\alpha\mu}(n,q)\right)\\
&+&\left.{n^\alpha\over 2Q\;n\cdot q} \bar
u(p_1)\left(\fmslash{\nb}\fmslash{p}_{1\perp}\Gamma P_{\bar n}+P_{\bar
n}\Gamma\fmslash{p}_{2\perp}\fmslash{n}\right)v(p_2)\right]\ep_\al(q)\deltascet(Q;p_1, p_2)+
\orderlambdasqr .\nonumber
\ea
The second line in \eqn{softexp} is reproduced in the effective theory by
$O_2^{(1b_n)}$, followed by emission of a soft gluon off the soft Wilson line
$Y_n$, and so has already been accounted for.  The term proportional to $\Delta^{\alpha\mu}(n,q)$ requires the
introduction of the operator
\ba O_2^{(1c_{ns})}(x)&=&-i\int_0^{\infty} dt \left[\bar\psi_n(\xnb)P_{\bar n} \Gamma
i\overleftarrow{D}^\mu_{\perp} \wnfun(\xnb, \xnbinf)\right]\\
&\times&\left[\ynfun(\xsninf,t n) i\overleftarrow{D}_{\perp\mu} (t n)
\ynfun(t n, 0)  \ynbfun(0, \xsnbinf)\right]\left[ \wnbfun(\xninf, \xn)P_\nb\psi_\nb(\xn)\right]\nonumber
\ea
while the higher order term in the multipole expansion of the momentum requires the operator
\ba
O^{(1\delta_{s})}_2(x)&=&Q\left[\bar\psi_n(\xnb)P_{\bar n} \Gamma\wnfun(\xnb,
\xnbinf)\right]\left[\ynfun(\xsninf, 0) (x_{\perp\mu} D_\perp^\mu+\overleftarrow{D}_\perp^\mu 
x_{\perp\mu})\ynbfun(0, \xsnbinf)\right]\nn
		&\times&\left[ \wnbfun(\xninf, \xn)P_\nb\psi_\nb(\xn)\right].
\ea
From \eqn{softexp} we find $C_2^{(1c_{ns})}=-2+\oas$ and $C_2^{(1\delta_{s})}=1+\oas$.

In addition to subleading corrections to the leading order amplitudes, at subleading order
additional processes -- soft quark and \ncol\ antiquark emission -- occur.
In the standard SCET approach, these arise from subleading terms in the effective
Lagrangian which directly couple the various sectors.  In our formulation, the only coupling
between the different sectors occurs via the external current $\J_2^{\rm SCET}$, so these processes 
are also described
in SCET by subleading operators $O_2^{(1i)}$.

\begin{figure}[t] 
   \begin{center}
   \subfigure[
\ncol\label{fig:softquark-n}]{\includegraphics[totalheight=0.12\textheight]
{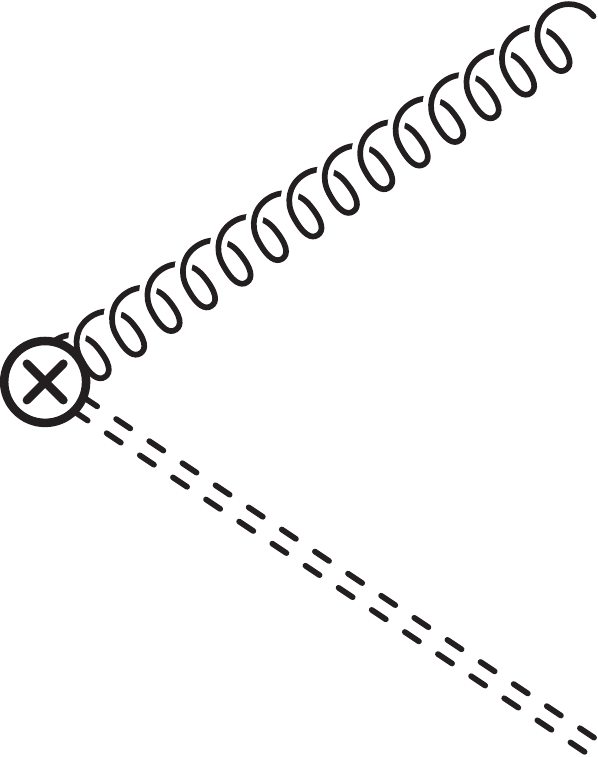}}\hspace{0.05\textheight}
   \subfigure[
soft\label{fig:softquark-s}]{\includegraphics[totalheight=0.12\textheight]
{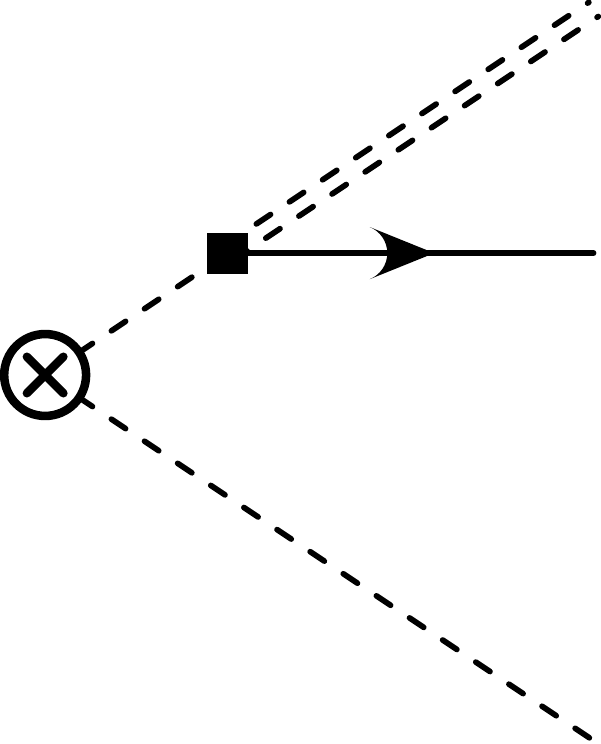}}\hspace{0.05\textheight}
   \subfigure[
\nbcol\label{fig:softquark-nb}]{\includegraphics[totalheight=0.12\textheight]
{diagrams/O2nb.pdf}}\hspace{0.05\textheight}   
   \caption{$O_2^{(1d_{ns})}$ as seen in each of the three sectors.  The single and double dashed
lines represent fundamental and adjoint Wilson lines
respectively.\label{fig:softquark}}
   \end{center}
\end{figure}

Consider \fig{fig:dijetqcd1} where the gluon is \ncol\ and the quark is soft.  
Expanding the amplitude \eqn{qcd1amp} for these kinematics gives
\ba\label{qcd2soft}
	i\amp_{as_q}&=&-ig
T^a\bar{u}_s(p_1)\frac{\Delta^{\al\mu}(\nb,q)}{n\cdot p_1}\frac{
\fmslash{n}}{2}\gm_\perp\Gamma P_{\bar n}v(p_2)\ep_\al(q)\deltascet(Q;q, p_2)+\orderlambdasqr .
\ea
The amplitude is reproduced in SCET by the subleading operator
\ba\label{o1sn}
	&&{O}^{(1d_{ns})}_2(x)=-i\int_0^\infty dt \left[\nb_\mu
igG_n^{a\mu\nu}(\xnb)\wnadj{}^{ab}(\xnb,\xnbinf)\right]\nn
	&&\times\left[\ynadj{}^{bc}(\xsninf,tn)\bar{\psi}_{s}(tn)
\ynfun(tn,0) T^c\frac{\fmslash{n}}{2}\gn_\perp\Gamma \ynbfun(0,\xsnbinf)
\right]\nn
	&&\times\left[\wnbfun(\xninf, \xn)P_\nb\psi_\nb(\xn)\right].
\ea
The structure of \eqn{o1sn} can be understood by generalizing the arguments that led to \eqn{o20}. 
The physical picture of \eqn{o1sn} is shown in \fig{fig:softquark}.  The \ncol\ gluon recoiling
against the soft quark and \nbcol\ antiquark in an $SU(3)$ adjoint looks like a gluon coupled to an
adjoint Wilson line, giving the first factor in \eqn{o1sn} and the picture \fig{fig:softquark-n}. 
The \nbcol\ sector sees no difference between an antiquark recoiling against an \ncol\ quark and
recoiling against an \ncol\ gluon and a soft quark in a relative fundamental state, so the third
factor is unchanged from \eqn{o20} and gives the picture \fig{fig:softquark-nb}.  Finally, the soft
sector has fundamental and anti-fundamental Wilson lines emitted by the current as usual, but then
the fundamental emits a soft quark and becomes an adjoint Wilson line (the \ncol\ gluon) as pictured
in \fig{fig:softquark-s}. From \eqn{qcd2soft} we find $C_2^{(1d_{ns})}=1+\oas$.

%%%%%%%%%%%%%%%%%%%%%%%%%%%
%%%%%%%%%%%%%%%%%%%%%%%%%%

\begin{figure}[t] 
   \begin{center}
   \subfigure[
\ncol\label{fig:tricolinear-n}]{\includegraphics[totalheight=0.12\textheight]
{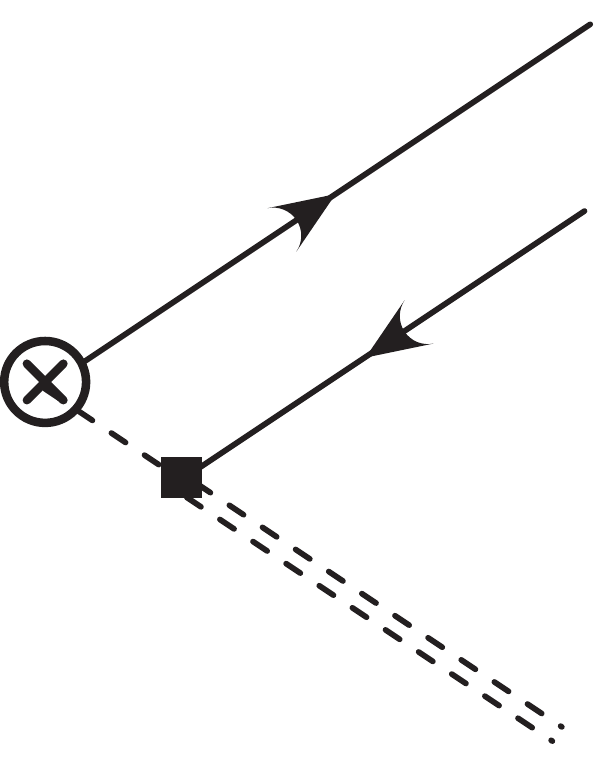}}\hspace{0.05\textheight}
   \subfigure[
soft\label{fig:tricolinear-s}]{\includegraphics[totalheight=0.12\textheight]
{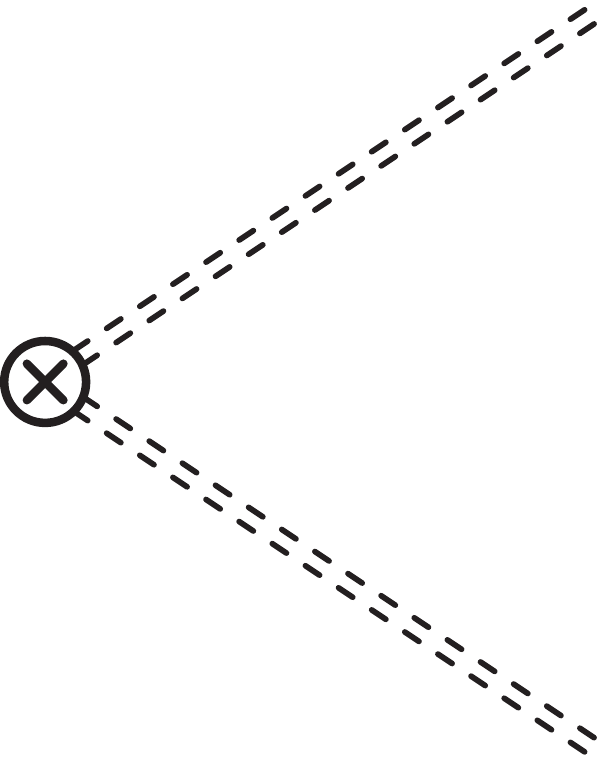}}\hspace{0.05\textheight}
   \subfigure[
\nbcol\label{fig:tricolinear-nb}]{\includegraphics[totalheight=0.12\textheight]
{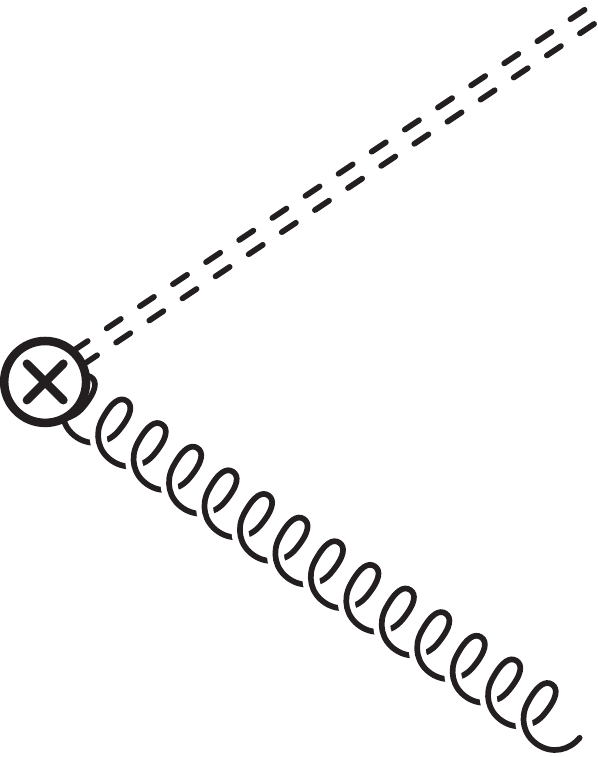}}\hspace{0.05\textheight}   
   \caption{$O_2^{(1f_{n n \bar n})}$ as seen in each of the three sectors.\label{fig:tricolinear}}
   \end{center}
\end{figure}

The situation is similar for emission of an \nbcol\ gluon recoiling against
an \ncol\ quark-antiquark pair.  Expanding the amplitudes \eqn{qcd1amp} and
\eqn{qcd2amp}, we find the leading order terms cancel between the two diagrams,
giving the amplitude
\ba\label{qcd2nn}
i\amp_{n n\bar n}&=&ig T^a\ep_\al(q) \bar u(p_1)\left({{\nbslash\over2}\ga_{\mu\perp}\Gamma P_n\over \nb\cdot
p_1}-{P_\nb\Gamma\ga_{\mu\perp}{\nbslash\over2}\over \nb\cdot p_2}\right)
v(p_2)\Delta^{\mu\alpha}(n,q)\deltascet(Q; p_1+p_2,q)\nn
	&&\qquad+\orderlambdasqr
\ea
These terms are reproduced in SCET by the operators
\ba
	O^{(1e_{n n \bar n})}_2(x)&=&-i\int_0^\infty dt
\left[\bar{\psi}_n(\xnb+t\nb)T^d\frac{\fmslash{\nb}}2 \gn_\perp\Gamma \wnfun(\xnb+t\nb, \xnb)P_n\psi_n(\xnb)
\wnadj{}^{dc}(\xnb+t\nb,\xnb^{\infty})\right]\nn
	&\times&\left[\ynadj{}^{c\hat b}(\xsninf, 0)\ynbadj{}^{\hat
bb}(0, \xsninf)\right]\left[n_\mu\wnbadj{}^{ba}(\xninf,\xn) ig
G_\nb^{a\mu\nu}(\xn)\right]
\ea
and
\ba
	O^{(1f_{n n \bar n})}_2(x)&=&-i\int_0^\infty dt \left[\bar{\psi}_n(\xnb)P_\nb\Gamma
\gn_\perp\frac{\fmslash{\nb}}2  \wnfun(\xnb, \xnb+t \nb)T^d\psi_n(\xnb+t\nb)\wnadj{}^{dc}
(\xnb+t\nb,\xnb^{\infty})\right]\nn
	&\times&\left[\ynadj{}^{c\hat b}(\xsninf, 0)\ynbadj{}^{\hat
bb}(0, \xsninf)\right]\left[n_\mu\wnbadj{}^{ba}(\xninf,\xn) ig
G_\nb^{a\mu\nu}(\xn)\right].
\ea
$O_2^{(1f_{n n \bar n})}$ is illustrated in the three frames in \fig{fig:tricolinear}.  
From \eqn{qcd2nn} we find $C_2^{(1e_{nn\nb})}=-C_2^{(1f_{nn\nb})}=-1+\oas$.
%%%%%%%%%%%%%%%%%%%%%%%%%%%
%%%%%%%%%%%%%%%%%%%%%%%%%%
%%%%%%%%%%%%%%%%%%%%%%%%%%
%%%%%%%%%%%

There are an additional four operators, defined analogously to the above operators, that arise due to 
corresponding corrections to the $\nb$ sector:
\ba
	O_2^{(1a_\nb)}(x)&=&\left[\bar\psi_n(\xnb)\frac{\nbslash}{2} \wnfun(\xnb,\xnbinf)\right]
	\left[\ynfun(\xsninf,0)\ynbfun(0,\xsnbinf)\right]\nn
		&&\times\left[\wnbfun(\xninf,\xn)i\overleftarrow{\fmslash{D}}_\perp(\xn)\Gamma P_\nb\psi_\nb(\xn)\right]
		\ea
		\ba
	O_2^{(1b_\nb)}(x)&=&\left[\bar\psi_n(\xnb)P_\nb\wnfun(\xnb,\xnbinf)\right]
	\left[\ynfun(\xsninf,0)\ynbfun(0,\xsnbinf)\right]\nn
		&&\times\left[\wnbfun(\xninf,\xn)\Gamma i\fmslash{D}_\perp(\xn)\frac{\nslash}{2}\psi_\nb(\xn)\right]
		\ea
		\ba
	O_2^{(1c_{\nb s})}(x)&=&-i\int_0^\infty dt\left[\bar\psi_n(\xnb)P_\nb\wnfun(\xnb,\xnbinf)\right]\nn
		&&\times\left[\ynfun(\xsninf,0)\ynbfun(0,t\nb)i{D}^{\mu}_\perp(t\nb)\ynbfun(t\nb,\xsnbinf)\right]\nn
		&&\times\left[\wnbfun(\xninf,\xn)iD_{\mu\perp}(\xn)\Gamma P_\nb\psi_\nb(\xn)\right]
		\ea
		\ba
	O_2^{(1d_{\nb s})}(x)&=&-i\int_0^\infty dt\left[\bar\psi_n(\xnb)P_\nb\wnfun(\xnb,\xnbinf)\right]\nn
		&&\times\left[\ynfun(\xsninf,0)\Gamma\gn_\perp\frac{\nbslash}{2}T^c 
		\ynbfun(0,t\nb)\psi_s(t\nb)\ynbadj{}^{cb}(t\nb,\xsnbinf)\right]\nn
		&&\times\left[\wnadj{}^{ba}(\xninf, \xn) igG_n^{a\nu\mu}(\xn)n_\mu\right].
\ea
These operators have the same matching coefficients as their $n$ sector counterparts. The one-gluon
Feynman rules for the operators $O_2^{(1i)}$ are shown in \fig{subleadingO2rules}.  Loop
calculations still require a zero-bin subtraction \cite{Manohar:2006nz}, which serves to fix the
double counting in the same way as the standard SCET.

Thus, we have shown the primary result of our paper for $e^+e^-\to$ dijet at $O(\al_s)$: $\orderlambda$ 
SCET effects can be written as QCD fields coupled to Wilson lines, where each sector is decoupled into 
a separate $SU(3)$ gauge theory.  The only expansion is in the current, and the subleading operators 
and physical pictures are generalizations of the $\orderlambdao$ operator \eqn{o20} and physical 
picture \fig{fig:o20}.

%===============B-->sgamma===================

\section{Heavy-to-Light Current}\label{sec:bsg}

A similar analysis may be carried out for $B\to X_s\gamma$ decay in the shape function region, 
\be
1-y\sim \Lambda_{\rm QCD}/m_b\sim \lambda^2
\ee
where $y=2 E_\gamma/m_b$ is the scaled energy of the photon.  In this region the light final-state
hadrons are constrained to form a jet, and SCET is the appropriate EFT.   The SCET analysis of this
process has been carried out to $O(\lambda^2)$ \cite{Lee:2004ja, Pirjol:2002km, Beneke:2004in,
Bosch:2004cb}.   In this section we present the operators up to $\orderlambda$ in order to show how
the picture introduced in this paper matches the standard SCET results.  The arguments are analogous
to the dijet analysis. However, now there is only one collinear sector and one soft sector, and a
copy of the Heavy Quark Effective Theory (HQET) Lagrangian is necessary.  Again, the collinear and
soft Lagrangian is not expanded in $\lambda$ and only the EFT current
\ba\label{bsgcurrent}
	\J^{\rm SCET}_h=e^{-i \frac{m_b}2( n + (1-y)\nb)\cdot x}\left[C_h^{(0)}O_h^{(0)}+\frac1{m_b}
	\sum_iC_h^{(1i)}O_h^{(1i)}+\orderlambdasqr\right] 
\ea
and the HQET Lagrangian are expanded in $\lambda$.  The phase in \eqn{bsgcurrent} corresponds to the 
removal of the large $b$-quark momentum, $m_b(n^\mu+\nb^\mu)/2$, and the outgoing photon 
momentum, $E_\gamma \nb^\mu$. 

%%%%%%%%%%%%%%%%%%%%%%%
\begin{figure}[tb]
   \begin{center}
   \subfigure[\label{fig:hqet1}]{\includegraphics[totalheight=0.07\textheight]
   {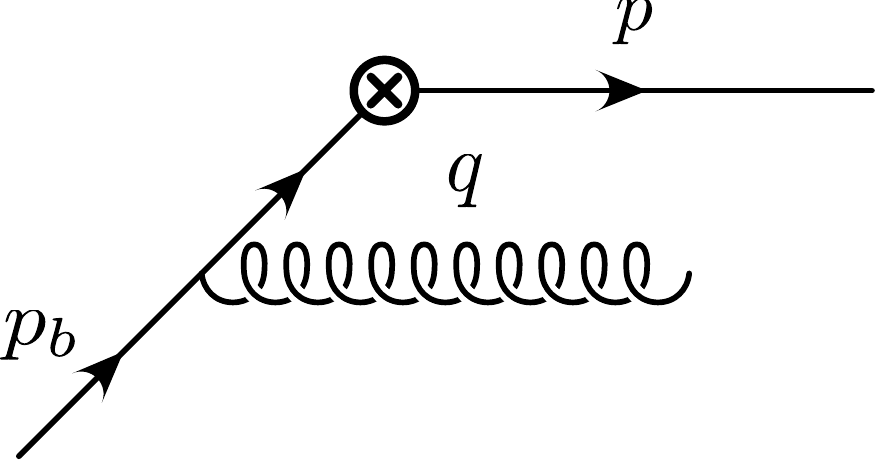}}\hspace{0.05\textheight}
   \subfigure[\label{fig:hqet2}]{\includegraphics[totalheight=0.07\textheight]
   {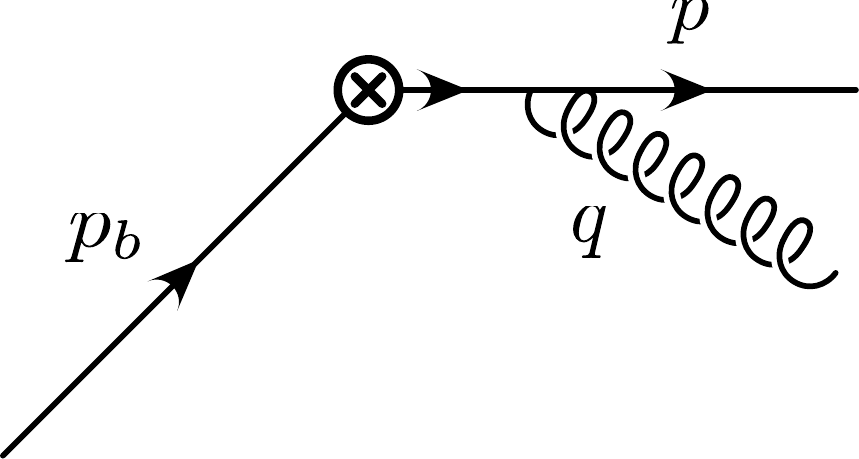}}
    \caption{One-gluon contributions to the matching of the QCD vertex for $B\to
X_s\gamma$. The quark, gluon, and heavy quark momentum are $p$, $q$, and
$p_b$ respectively.
\label{fig:hqet}}
   \end{center}
\end{figure}
%%%%%%%%%%%%%%%%%%%%%%%

The relevant $\orderalpha$ graphs are in \fig{fig:hqet} and have amplitudes 
\ba\label{bsga}
	i\amp_a^B=igT^a\bar u(p)\Gamma\frac{2p_b^\al -\fmslash{q}\ga^\al}{2p_b\cdot q} u(p_b)\ep_\al(q)
	\deltaqcdbsg(m_b, y; p+q+p_b)
\ea
and
\ba\label{bsgb}
	i\amp_b^B=-igT^a\bar u(p)\frac{2p^\al+\ga^\al\fmslash{q}}{2p\cdot q}\Gamma u(p_b)\ep_\al(q)
	\deltaqcdbsg(m_b, y; p+q+p_b).
\ea
in QCD.  The amplitude expansions are similar to the dijet case.  The collinear and heavy quark
spinors are expanded using the Dirac Equation. The collinear quark is done in \eqn{spinorexp} and
the heavy quark expansion is $u(p_b)=\left( 1+\orderlambdasqr\right)h_v(k)$where $h_v$ is an HQET
heavy quark field satisfying $(\nslash+\nbslash)h_v=2h_v$.  As before, the QCD momentum conserving
$\delta$ function, 
\ba
	\deltaqcdbsg(m_b, y; p)&=&2\,\delta\left(m_b-p\cdot \nb)\right)\delta\left(m_b(1-y)-p\cdot n\right)
	\delta^{(2)}\left(\vec p_\perp\right),
\ea
is expanded onto the SCET momentum conserving $\delta$ function,
\ba\label{bsgmmmatch}
	\deltaqcdbsg(m_b, y; p)=\deltascet(m_b,y;p_n,p_s)+\ldots,
\ea
where
\ba\label{bsgmm}
	\deltascet(m_b,y;p_n,p_s)&=&2\,\delta(m_b-p_n\cdot\nb)\delta(m_b(1-y)-(p_n+p_s)\cdot
n)\delta^{(2)}(\vec{p}_{n,\perp})
\ea
and the higher moments are reproduced by higher orders of the SCET current \eqn{bsgcurrent}. The
residual $b$ quark momentum $k^\mu\sim\lambda^2m_b$ is included in $p_s$ with the appropriate sign. 
Unlike in the dijet case, all components of the collinear momentum and the $p_s\cdot n$ component of
the soft momentum are constrained. The leading order expansion of \eqn{bsga} and \eqn{bsgb} are
reproduced by the operator
\ba\label{Oh0}
	O^{(0)}_h(x)&=&\left[\bar{\psi}_n(\xc)P_\nb \Gamma\,\wnfun(\xc,
\xinftyc)\right]\left[ \ynfun(\xshinf, \xsh) h_v(\xsh)\right]
\ea
with $C_h^{(0)}=1+\orderalpha$, which is similar to the leading order label SCET operator \cite{Bauer:2000ew}
\be \label{usualbsg}
	\tilde \J^{(0)}_{h}=e^{-i \frac{m_b}2( n + (1-y)\nb)\cdot x}\,\bar \xi^{(0)}_{n, \tilde p}(x) 
	\wnfun(x,\infty) \Gamma \ynfun(\infty,x) h_v(x).
\ee
The difference between \eqn{Oh0} and \eqn{usualbsg} are the use of four-component spinors for the 
collinear fields and the fields are at the positions
\ba\label{bsgx}
	\xc&=(x\cdot n,x\cdot \nb,x^\perp) \quad \xsh&=(0,x\cdot \nb, 0) \nn
	\xinftyc&=(-\infty, x\cdot\nb, x^\perp) \quad \xinftysh&=(0,\infty,0).
\ea
The positions are chosen to reproduce \eqn{bsgmm}.

The $\orderlambda$ expansion of the amplitudes is done in the same way as in the previous section, so 
we omit the details.  We find the following subleading operators:
\ba\label{bsgnlo}
	O_h^{(1a_c)}(x)&=&\left[\bar\psi_n(x)P_\nb\Gamma i\fmslash{D}_\perp(x)\frac{\fmslash{n}}2 
	\wnfun(x,\xinftyc)\right]\left[\ynfun(\xinftysh,x_s)h_v(x_s)\right]\nn
	O_h^{(1b_c)}(x)&=&\left[\bar\psi_n(x)\frac{\nbslash}{2}i\overleftarrow{\fmslash{D}}_\perp(x)\Gamma 
	\wnfun(x,\xinftyc)\right]\left[\ynfun(\xinftysh,x_s)h_v(x_s)\right]\nn
	O_h^{(1c_s)}(x)&=&-i\int_{-\infty}^0 dt\left[\bar\psi_n(x)P_\nb i\overleftarrow{D}_{\perp\mu}(x)
	\Gamma\wnfun(x,\xinftyc)\right]\\
		&&\times\left[\ynfun(\xinftysh, x_s+tn)i\overleftarrow{D}^\mu_\perp(x_s+tn)\ynfun(x_s+tn,x_s)h_v(x_s)\right]\nn
	O_h^{(1d_s)}(x)&=&-i\int_{-\infty}^0 dt\left[\nb_\mu igG^{a\mu\nu}_n(x)\wnadj{}^{ab}(x,\xinftyc)\right]\nn
		&&\times\left[\ynadj{}^{bc}(\xinftysh, x_s+tn)\bar\psi_s(x_s+tn)T^c\frac{\fmslash{n}}2\gn_\perp
		\Gamma\ynfun(x_s+tn,x_s)h_v(x_s)\right]\nn
	O_h^{(1\delta_s)}(x)&=&m_b\left[\bar\psi_n(x)P_{\bar n} \Gamma\wnfun(x,\xinftyc)\right]
	\left[\ynfun(\xinftysh, x_s)
	\left( \overleftarrow{D}^\mu_\perp x_{\perp\mu}+x_{\perp\mu}\overrightarrow{D}_\perp\right)(x_s) h_v(x_s)\right].\nonumber
\ea
These operators are the analogous dijet operators with only one collinear sector. The operators
$O_h^{(1c_s,1s)}$ have integration limits $-\infty$ and $0$ since the $b$-quark is coming in from
$-\infty$ (as opposed to the dijet case where the partons are outgoing to $+\infty$).  The matching
coefficients are $C_h^{(1a_c, 1b_c,1\delta_s)}=1+\orderalpha$, $C_h^{(1d_s)}=-1 + \orderalpha$ and
$C_h^{(1c_s)}=2+\orderalpha$.  

The formulation of SCET introduced in this paper must be equivalent to standard SCET formulations at
subleading orders.  For example, the soft quark operator $O_h^{(1d_s)}$ is reproduced by the
time-ordered product of $\lag_{\xi q}^{(1)}$ and the leading order current $\tilde\J_h^{(0)}$ of
\eqn{usualdijetcurrent}.  A particularly simple example of a subleading standard SCET operator that
can be re-written as one of  our operators is the $\orderlambdasqr$ label SCET current
\cite{Beneke:2002ph, Beneke:2002ni} 
\ba \label{usualnnlocorrection}
	J^{(2A)}=\bar\xi(x)\frac{1}{i\nb\cdot D}\Big[in\cdot D(x) W_c(x,\xinftyc)\Big] h_v(x_s) +\ldots \,,
\ea
which is identical to the operator we find at $\orderlambdasqr$
\ba\label{Oh2ac}
	O_h^{(2a_c)}&=&-i\, m_b\int_{-\infty}^0 dt\left[\bar\psi_n(x)P_\nb\wnfun(x,x+t\nb) in\cdot D(x+t\nb)
	\Gamma\wnfun(x+t\nb, \xinftyc)\right]\nn &&\times\left[\ynfun(\xinftysh,x_s)h_v(x_s)\right]
\ea
with $C_h^{(2a_c)}=1+\orderalpha$. The ``$\ldots$'' in \eqn{usualnnlocorrection} refer to other 
$\orderlambdasqr$ terms. The equivalence between \eqn{usualnnlocorrection} and \eqn{Oh2ac} can be 
shown using the  relations \cite{Bauer:2001ct, Beneke:2002ph} 
\be\label{inverseD}
	\frac1{i\nb\cdot D} \wnfun(x,\xinftyc) =\wnfun(x,\xinftyc) \frac1{i\nb\cdot\partial}
\ee
and
\be\label{inversed}
	\frac{1}{i\nb\cdot\partial}\phi(x)=-i\int_{-\infty}^0 dt\;\phi(x+t\nb),
\ee
and the field redefinition \eqn{redef}.

%===============Conclusion============
\section{\label{sec:conc}Conclusions}

We have demonstrated how SCET can be written as a theory of separate, decoupled sectors of QCD by
explicitly performing the matching of the external current at tree level in $\al_s$ to subleading
order in $\lambda$ for both dijet production and $B\to X_s\gamma$. ÊInteractions between different
sectors are reproduced in SCET by Wilson lines. ÊWe believe this makes the SCET picture more
transparent:  instead of a complicated collinear Lagrangian that couples two-component collinear
quarks to soft fields, the Lagrangian is just multiple QCD copies. The only expansion in $\lambda$
occurs in the currents, which are QCD fields coupled to  Wilson lines that represent the colour flow
of the other sectors. ÊThe subleading currents are generalizations of the leading order currents
akin to higher twist corrections to light-cone distribution functions. ÊCorrections to leading-order
factorizations theorems should be simpler since the manifest decoupling of sectors that occurred at
leading order now exists to all orders.

%===============================

\begin{acknowledgments}
This work was supported by the Natural Sciences and Engineering Research Council
of Canada. We would like to thank C. Bauer, W. Cheung and Z. Ligeti for discussions.
\end{acknowledgments}

\bibliography{bibliography}

\end{document}